\documentstyle[12pt,psfig]{article}
\normalsize

\let\oldtheequation=\theequation
\def\doteqs#1{\setcounter{equation}{0}
            \def\theequation{{#1}.\oldtheequation}}

\newcounter{sxn}
\def\sx#1{\addtocounter{sxn}{1} \bigskip\medskip \goodbreak
\noindent{\large\bf
\centerline{\thesxn.~~#1}} \nobreak \medskip}
\def\sxn#1{\sx{#1} \doteqs{\thesxn}}

\newcounter{axn}

\def\br{\begin{thebibliography}{abc}}
\def\er{\end{thebibliography}}

\date{}

\begin{document}

\textheight=22 true cm
\textwidth=15 true cm
\normalbaselineskip=24 true pt
\normalbaselines
\topmargin -0.5 true in
\bibliographystyle{unsrt}
\def\sl{\em}
\def\be{\begin{equation}}
\def\ee{\end{equation}}
\def\bea{\begin{eqnarray}}
\def\eea{\end{eqnarray}}

\setcounter{page}{0}
\thispagestyle{empty}

\begin{flushright}
{\sf SINP-TNP/98-04}
\end{flushright}

\begin{center}
{\Large \bf Dispersive Bounds on The Shape Of \\
${\unboldmath \Lambda_b \to \Lambda_c l {\bar \nu_l}}$ Formfactors}\\[5mm]
{\large\sf Debrupa Chakraverty\footnote{Electronic address:
 rupa@tnp.saha.ernet.in}, Triptesh De\footnote{Electronic address:
 td@tnp.saha.ernet.in}, Binayak Dutta-Roy\footnote{Electronic
 address: bnyk@tnpdec.saha.ernet.in}}\\
and\\[5mm]
{\large\sf K. S. Gupta\footnote{Electronic address:
gupta@tnp.saha.ernet.in}}\\
{\sl Saha Institute of Nuclear Physics,\\
1/AF Bidhannagar, Calcutta - 700 064, India}\\[5mm]
\end{center}

\begin{abstract}
We derive a theoretically allowed domain for the  charge radius
$\rho$ and curvature $c$ of the Isgur-Wise function
describing the decay $\Lambda_b \to \Lambda_c l {\bar \nu_l}$.
Our method uses crossing symmetry, dispersion relations and analyticity 
in the context of the Heavy Quark Effective Theory but is independent 
of the specifics of any given model. The experimentally determined 
values of the $\Upsilon$ masses have been used as input information. 
The results are of interest for testing different models employed to 
calculate the heavy baryon formfactors which are used for the extraction 
of $\vert V_{cb} \vert$ from the experimental data. 
\end{abstract}
\bigskip

\newpage
\sxn{INTRODUCTION}
 During the last few years,   considerable 
 progress  has been made in the development of methodologies for 
 the extraction of $\vert V_{cb} \vert$ 
[1-6] from  both 
  exclusive and inclusive semileptonic B meson decays, where the leptonic 
current is clearly separated from the matrix elements of the 
hadronic current.
 The CKM matrix element $V_{cb}$ has 
important implications for the investigation
 of rare decays and CP violation.  It is therefore imperative to have as many
 independent and accurate determinations of $\vert V_{cb} \vert$ as possible. 
 There are however some theoretical limitations in the determination of
 $\vert V_{cb} \vert$ from inclusive semileptonic decays due to  issues 
 related to  validity of quark-hadron duality near the kinematic endpoint 
region \cite{boyd1}. On the other hand, the exclusive decays must be described
 in terms of a number of  undetermined nonperturbative formfactors 
 that contain the
 physics of the hadronisation process.

From the theoretical point of view, in the limit where the heavy quark mass
tends to infinity,  the analysis of  
 heavy baryonic semileptonic decays are comparable to that of 
 semileptonic decays of heavy mesons. This is due to the fact that
  heavy quark symmetry in the above limit predicts  a
 single universal formfactor for both these types of decays known as the
Isgur-Wise (IW) function \cite{iw1}.
It is therefore interesting to estimate  $V_{cb}$ from the decays of heavy
baryons.
However,  experimental data on the decay of heavy baryons is still
 sparse  compared to the corresponding meson decays. Semileptonic decays of the
 $\Lambda_b$ baryon have up to now only been observed at LEP \cite {pdg,op}.
It is however expected that in the near future more  data would be available 
from the LEP  as well as from the the forthcoming B 
factory. Once these results are  available, the theoretical value 
for the semileptonic decay width of $\Lambda_b$ can be used to gain
 an independent determination of $\vert V_{cb} \vert$.

 In  semileptonic decays of both heavy mesons and baryons, the heavy quark
symmetry predicts the value of the corresponding IW functions at zero
recoil \cite{iw1, neu1},  though the shapes of the IW functions are 
left unspecified 
and require detailed knowledge of the non-perturbative
strong interaction physics. Unfortunately, the differential decay width 
${d\Gamma \over {d q^2}}$ vanishes at zero recoil and hence the
 known normalization of this function at that point is not enough to 
extract  $V_{cb}$ 
from the data.
To overcome this difficulty various models are used to determine the
functional form of the IW function and  $V_{cb}$ is extracted therefrom.
However,  the observed variations of $\vert V_{cb} \vert $ obtained from
 different models turn out to be  larger than the statistical and 
 systematic uncertainties in the experiments.
 From the experimentally  
measured lepton invariant mass spectrum
  one may determine $V_{cb}$ in a model 
independent way by extrapolating to
 the kinematical endpoint of maximal 
momentum transfer to the leptons corresponding to
 the zero recoil point. 
However, this extrapolation requires a large amount of data, very close to
 $q_{max}^2$, which is difficult to assess experimentally. 
Thus measurement
 of $\vert V_{cb} \vert $ requires a parametrization  of the
 IW function. It is therefore necessary to put constraints as 
 far as possible in a model-independent manner on this function.

	In this paper we present an analysis of the model independent
constraints on  heavy baryonic IW function arising in the description  of 
semileptonic ${\Lambda_b \to \Lambda_c l {\bar \nu_l}}$ decay.
 This is accomplished by first obtaining  constraints on the charge radius and
the convexity of the IW function for the elastic $\Lambda_b$  formfactors 
 using
the techniques of dispersion relations and analyticity. Next,
 heavy quark symmetry
is used to relate this IW function to the formfactors appearing in the 
semileptonic ${\Lambda_b \to \Lambda_c l {\bar \nu_l}}$ decay. 
 The dispersion theoretic 
method of extracting information on amplitudes is quite 
old \cite{o1,o2,o3} and
 has  been applied to the 
study of semileptonic decays of light mesons in a more contemporary
  language \cite{l1,l2}. Its application to 
heavy quark systems  has received much attention in more recent 
years [7, 17 - 24]. In Section 2 we give a brief account of
 the general formalism
for the decay of heavy baryons. The method for the estimation of model
independent bounds on the baryonic IW function is described in Section 3.
The numerical results are presented in Section 4,
 and Section 5 concludes the paper with some discussions and future outlook. 

\sxn{GENERAL FORMALISM OF HEAVY BARYON DECAY}
	In this section we give a brief account of the semileptonic 
decay of  $\Lambda_b$ baryon using heavy quark symmetry \cite{cheng}.
 
The hadronic matrix element for the
  $\Lambda_b \to \Lambda_c l {\bar \nu_l}$ decay is parametrized by the
 following general decomposition consistent with Lorentz invariance.
\bea
 \langle \Lambda_c (p^{\prime}) \vert J_{\mu} \vert
 \Lambda_b (p) \rangle & = & {{\bar u}_{\Lambda_c}}\{ [f_1(q^2)\gamma_{\mu}
+i f_2(q^2)\sigma_{\mu \nu} q^{\nu}+f_3(q^2)q_{\mu}]\nonumber\\
& &  -[g_1(q^2)\gamma_{\mu}
+i g_2(q^2)\sigma_{\mu \nu} q^{\nu}+g_3(q^2)q_{\mu}]\gamma_5\}u_{\Lambda_b},
\eea
where $J_{\mu}= (V_{\mu} - A_{\mu})={\bar c}\gamma_{\mu} (1-\gamma_5)b$
 and $q= (p - {p^\prime})$ is the momentum transferred to the leptons.
Here  $u_{\Lambda_b}$ and  $u_{\Lambda_c}$ are the Dirac 
spinors of $\Lambda_b$
 and $\Lambda_c$ respectively. The $f'$s and $g'$s are 
the formfactors for the independent 
 Lorentz covariants corresponding to the hadronic matrix elements of
 the vector and axial vector currents respectively.

 When both baryons are heavy, 
it is convenient to parametrize the matrix element
 in terms of the four velocities $v$ and $v^{\prime}$ \cite{cheng} :
\bea
 \langle \Lambda_c (v^{\prime}) \vert J_{\mu} \vert
 \Lambda_b (v) \rangle & = & {{\bar u}_{\Lambda_c}}\{
F_1(\omega)\gamma_{\mu}
+ F_2(\omega)v_{\mu}+F_3(\omega){v^{\prime}}_{\mu}\nonumber\\
& &  -[G_1(\omega)\gamma_{\mu}
+ G_2(\omega) v_{\mu}+G_3(\omega){v^{\prime}}_{\mu}]\gamma_5\}u_{\Lambda_b},
\eea
 where the formfactors [written with corresponding uppercase $F'$s and $G'$s]
 are functions of 
  $\omega= v \cdot v^{\prime}$,  the only non-trivial Lorentz scalar 
 formed from $v$ and $v^{\prime}$.

 In terms of these formfactors, 
the differential decay width in the zero mass approximation 
for charged lepton
 is given by
\be
{d\Gamma \over {d q^2}}={{\vert V_{cb} \vert^2 {G_F}^2  k t^{1\over 2}}\over
 {96 \pi^3 m_{\Lambda_b}^3}}\{t_- [2 t \vert F_1 \vert^2 + \vert H_V
\vert^2]
+ t_+ [2 t \vert G_1 \vert^2 + \vert H_A \vert^2]\},
\ee
where
\be
H_V(q^2)= (m_{\Lambda_b} + m_{\Lambda_c})F_1 +{t_+\over 2} ({F_2 \over
 m_{\Lambda_b}} + {F_3 \over m_{\Lambda_c}}),
\ee
\be
H_A(q^2)= (m_{\Lambda_b} - m_{\Lambda_c})G_1 -{t_-\over 2} ({G_2 \over
 m_{\Lambda_b}} + {G_3 \over m_{\Lambda_c}}),
\ee
 with
$t=q^2,$
$t_\pm = (m_{\Lambda_b} \pm m_{\Lambda_c})^2 - t,$
and $ k= {1\over 2} \sqrt{{t_+ t_-}\over t}.$

In the infinite quark mass limit, the spin-flavour symmetry for the
 b and c quark relates all the formfactors to a single universal function
 $\xi(\omega)$, which is the IW function.
$$F_1(\omega)= G_1(\omega)=\xi(\omega),$$
and 
$$F_2(\omega)= F_3(\omega)= G_2(\omega)= G_3(\omega)=0.$$
 The IW function so defined  is independent of heavy quark
masses and normalized at zero recoil 
 $(v=v^{\prime})$ i.e. $\xi(1)=1$.

The behaviour of the 
IW function close to zero recoil is of  particular
 interest and it is customary to parametrize this behaviour 
through  a Taylor's  
series expansion
 about the zero recoil point namely,
 \be
\xi(\omega)= 
1 - \rho^2 (\omega -1) + c (\omega -1)^2 + {\cal{O}}[(\omega-1)^3],
\ee
where $\rho$ and $c$ are  
 referred to as the charge radius and the convexity  parameter respectively. 
 The kinematic region, accessible to the semileptonic decay lies in
such a small domain ($ w = 1$ to $1.43$) that a precise knowledge of 
the charge radius could basically determine the IW
 function in the physical region and the convexity parameter 
gives the corrections to
the simple linear dependence.
In the following section we derive  bounds on  
$\rho$ and $c$ and thereby provide  model independent constraints on the IW
function.

\sxn{DERIVATION OF THE BOUNDS}
 We consider the elastic formfactors of the $\Lambda_b$
 baryon through the matrix elements of the flavour-conserving 
 current $V_{\mu} = {\bar b} {\gamma_\mu} b$
\be
 \langle \Lambda_b (p^{\prime}) \vert V_{\mu} \vert 
 \Lambda_b (p) \rangle  =  
{{\bar u}_{\Lambda_b}}\{ F^E_1(\omega)\gamma_{\mu}
+ F^E_2(\omega)(v+{v^{\prime}})_{\mu}
\}u_{\Lambda_b}.
\ee
In the heavy quark limit, the b-number conserving elastic formfactors are 
 related to the same IW function $\xi(\omega)$,
 which enters in the expression for the 
differential decay rate  $\Lambda_b \to \Lambda_c l {\bar \nu}$.
The short distance and the finite mass 
corrections are in this case much smaller than that for the formfactors 
 involving  flavour changing currents, corresponding to
 the decay $\Lambda_b \to \Lambda_c l {\bar \nu}$.
 
 The method is based on the properties of 
the two-point  function $\Pi(q^2)$
 defined as
 \bea
\Pi^{\mu \nu}(q^2) & = & 
i \int d^4 x e^{i q\cdot x}\langle 0 \vert T(V^{\mu}(x) V^{\nu}(0)\vert 0 
\rangle\nonumber\\
 & = & (q^{\mu} q^{\nu} - q^2 g^{\mu \nu}) \Pi(q^2),
\eea
with $V^{\mu}={\bar b}{\gamma^\mu} b$.
The conservation of the current $V^{\mu}$ leads to the 
transverse nature of the two-point function.
 In QCD, the asymptotic nature of the two-point function $\Pi(q^2)$ is
 such that it satisfies a once subtracted dispersion relation
\be
\chi(Q^2)={{\partial \Pi(q^2)} \over {\partial q^2}}{\vert}_{q^2 = - Q^2} =
 {1\over \pi} \int_0^\infty {{\rm Im} \Pi(t) \over {(t + Q^2)^2}} dt.
\ee
The absorptive part ${\rm Im} \Pi(t)$ is obtained from the unitarity relation
\be
 (q^{\mu} q^{\nu} - q^2 g^{\mu \nu}) {\rm Im} \Pi(q^2)= 
{1\over 2}\sum_\Gamma d \mu_{\Gamma}
 (2 \pi)^4 \delta^{(4)}(q -p_{\Gamma})
\langle 0 \vert (V^{\mu}(0) \vert \Gamma 
\rangle \langle \Gamma \vert V^{\nu}(0)\vert 0 \rangle,
\ee
where the summation is extended  over all possible
 intermediate  hadronic states with the quantum numbers
 of the $V^{\mu}$ current and is  weighted by the corresponding 
  phase space factor $d\mu_{\Gamma}$. A judicious choice of $\mu$
 and $\nu$ $(\mu=\nu)$ makes this a sum of positive
definite terms and thus  one can obtain 
 a strict inequality by retaining  only
 the term with intermediate  
 $\Gamma =\Lambda_b {\bar \Lambda_b}$ state. From crossing symmetry, the
 $\Lambda_b {\bar \Lambda_b} \to$ vacuum matrix 
element is described by the same set of 
  elastic formfactors, but relevant to the pair-production region 
$( 4 m_{\Lambda_b}^2 \leq q^2\leq \infty)$
 instead of the elastic region $(q^2
 \leq 0)$.
 This gives rise to an integral inequality
\be
\chi(Q^2)\geq {1\over {24 \pi^2}}\int_{4 m_{\Lambda_b}^2}^{\infty} 
{dt\over {(t+Q^2)^2}}
\sqrt{1 -{4 m_{\Lambda_b}^2\over t}}[ 2 t \vert F_1^E \vert^2 + \vert 
H_V^E\vert^2],
\ee
where
\be
H_V^E(q^2)= 2m_{\Lambda_b} F_1^E +(4  m_{\Lambda_b}^2 - t) 
 {F_2^E \over m_{\Lambda_b}}.
\ee
 In the case of mesons, inclusion of other two-particle intermediate states
 related by spin-flavour symmetry were made
 to saturate the dispersion relation in order to find the relevant bound.
 For $\Lambda_b$
 baryon decay, there are 
 no spin symmetric partners that contribute
 to the dispersion relation bound. However,
 the presence of more than one helicity amplitude fulfills in effect 
the same role.

For $Q^2$ far from the resonance region 
$ (2 m_b \Lambda_{QCD} \ll  4 m_b^2 + Q^2)$,
 the  two-point function can be computed 
reliably from perturbative QCD. For large b quark mass,
 it is sufficient to take $Q^2=0$.
The one loop expression for $\chi(0)$ is
 \be
\chi(0) = {3\over 2 \pi^2} \int_0^1 {{x^2 (1 - x^2)}\over m_b^2} dx.
\ee
Perturbative $\alpha_s$ corrections to this
 result are negligibly small at the physical 
$m_{\Lambda_b}^2$ scale. The non-perturbative
 corrections \cite{svz} are included  by
 expressing the two-point function as an
  operator product expansion  and incorporating the leading nonperturbative
  gluonic and quark-antiquark vacuum condensate terms $\langle G^2\rangle$
 and $\langle q {\bar q}\rangle$. But these terms
 are suppressed by  the fourth 
 powers of a large mass scale for dimensional reasons.

The analysis simplifies by using a conformal transformation 
 to map the full complex $t$ plane onto the unit disc in the complex $z$ plane,
\be
{{1 +z}\over {1-z}}= \sqrt{1 -{t\over {4 m_{\Lambda_b}^2}}}.
\ee
The  transformation  maps the cut $[4{m_{\Lambda_b}^2}, \infty]$ on the unit 
$z$ circle and the rest of the $t$ plane
 into the open unit $z$ disc. In terms of the new variables $z$  the 
inequality  eq. (3.5) gets translated to
\be
{1\over {2 \pi i}} \oint_C {dz\over z}
 [ \vert {\phi_{F_1^E}}(z) F_1^E(z)\vert^2
 +  \vert {\phi_{H_V^E}}(z) H_V^E(z)\vert^2]\leq 1,
\ee
where  the functions $\phi_{F_1^E}$ and $\phi_{H_V^E}$
are defined as 
\be
 \phi_{F_1^E} = \sqrt{{5 \pi}\over 96} (1 + z) (1-z)^{1/2},
\ee
and
\be
 \phi_{H_V^E} = {1\over 32} \sqrt{{5 \pi}\over 3} (1 + z)
 (1-z)^{3/2} {m_{\Lambda_b}^{-1}}.
\ee
These functions are analytic and nonzero inside the unit disc 
$\vert z\vert < 1$.
 Their moduli squared on the boundary are equal to the
 positive weights appearing
 in the integrals, multiplied by the Jacobian $\vert {dt\over dz}
 \vert$ of the
 conformal transformation.
 To translate the bound into one that is valid in the semileptonic
 region, we need a
 function which is analytic inside the unit disc. The functions
 $\phi_i$ have
 no poles, branch cuts
 or zeros in the interior of the unit circle $\vert z \vert <1$,
 but the
 formfactors $F^E(q^2)$  and $H_V^E(q^2)$ have poles due to
 the existence of 
spin-one $b{\bar b}$ states $(J^P=1^-)$ below the
 $\Lambda_b{\bar \Lambda_b}$ threshold. They  also have branch cuts
 originating 
 from
 non-resonant  continuum contributions with the invariant masses below
 $\Lambda_b{\bar \Lambda_b}$ threshold, though their effects
 may be neglected \cite{boyd2, boyd4}. The relevant 
states giving rise to poles in this context   are the $~^3S_1$ and $~^3D_1$ 
bottomonium states,
 i.e. $\Upsilon(nS)$ and $\Upsilon(nD)$, respectively. From the experimental 
 values of the masses of the
 $\Upsilon$ resonances, the position of the poles are obtained at the points:
\be
 z_1= -0.29,~ z_2=-0.37,~z_3=-0.43,~z_4= -0.48,~z_5=-0.58,~z_6=-0.65.
\ee
The residues of the formfactors at these poles are however unknown. To 
make up for this lack
 of information, one introduces  a product of
 the functions of the form $(z-z_i)/(1-{\bar z_i} z)$,
 known as Blaschke factors \cite{duren}
\be
P(z) = \prod_{j=1}^6 {{(z-z_j)}\over {(1-{\bar z_j} z)}}.
\ee
This function $P(z)$ has the virtue that it is analytic on the unit disc
 $\vert z_i \vert<1$, and eliminates, through multiplication,
 the  poles of $F_i^E$
 at each $z=z_j$. Since each term in $P(z)$ is contrived to
 be unimodular on the 
boundary, the inequality
 remains unchanged. Then the inequality (3.9) becomes
\be
{1\over {2 \pi i}} \oint_C {dz\over z}
 [ \vert {\phi_{F_1^E}}(z) P(z) F_1^E(z)\vert^2
 +  \vert {\phi_{H_V^E}}(z)P(z) H_V^E(z)\vert^2]\leq 1.
\ee
It is important to note that up to now the formfactors are treated as 
 completely unknown functions, 
 because heavy quark symmetry is quite inapplicable
 in the timelike domain near the
 pair-production threshold \cite{ball},
 though  is only valid in the neighbourhood of  the zero recoil point. 

 Both the functions $\phi_i(z)$ and $P(z)F_i^E(z)$
 are now analytic in the unit
 disc. We can hence apply the well known results of interpolation
 theory \cite{duren} for vector-valued
 analytic functions to obtain the bounds on the formfactors
 at points inside the unit circle.
Accordingly, we apply  the inequality of the Schur-Caratheodory type
 \cite{duren} at the origin  retaining
 terms up to the first
 derivative with respect to $z$
\bea
 \phi_{F_1^E}^2(0) P^2(0) F_1^{2E}(0)
& + & \phi_{H_V^E}^2(0) P^2(0) H_V^{2E}(0)
+ ( \phi_{F_1^E} P F_1^E)_z^{\prime 2}(0)\nonumber\\
& + & ( \phi_{H_V^E} P {H_V^E})_z^{\prime 2}(0) \leq 1,
\eea
where the prime denotes differentiation with respect to $z$.
Through the conformal mapping, the point 
 $\omega=1$ is mapped to $z=0$.  We can hence 
 use the prediction of 
heavy quark symmetry for the formfactors at the point of zero recoil,
 which leads to
\be
F_1^E(z=0)=\xi(z=0)=1, ~~~ H_V^E(z=0)=2 m_{\Lambda_b}
 \xi(z=0)= 2 m_{\Lambda_b}.
\ee
The charge radius and convexity of the IW function are defined by
\be
\rho^2=-\big [{{d\xi(\omega)} \over {d \omega}} \big ]_{\omega=1},
~~~c==-\big [{ {d^2 \xi(\omega)} \over {d \omega^2}} \big ]_{\omega=1},
\ee
the derivatives here being  with respect to $\omega$.
 They are related to 
 derivatives with respect to $z$ 
through
\be
\big [{ {d\xi(\omega)} \over {d z}} \big ]_{z=0}
=-8 \rho^2,~~~ 
\big [{ {d^2\xi(\omega)} \over {d z^2}} \big ]_{z=0}
= 128 c - 32 \rho^2.
\ee
From Eqs. (3.15 - 3.18),  the upper and lower bounds on the charge radius 
are readily obtained 
 to read
\be
{{b - \sqrt{b^2 - 4 a d}}\over a} \leq \rho^2 
\leq {{b + \sqrt{b^2 - 4 a d}}\over a}
\ee
with
\be
a = 64 P^2(0)[\phi_{F_1^E}^2 + \phi_{H_V^E}^2 ],
\ee
\be
b = 8 P(0) [\phi_{F_1^E}(0) (\phi_{F_1^E} P)^{\prime}_z(0)+
\phi_{H_V^E}(0) (\phi_{H_V^E} P)^{\prime}_z(0)],
\ee
and
\be
d= 
 P^2(0)[\phi_{F_1^E}^2 + \phi_{H_V^E}^2 ]+
(\phi_{F_1^E} P)^{\prime 2}_z(0)
+(\phi_{H_V^E} P)^{\prime 2}_z(0) -1.
\ee
 When terms up to the second derivatives are included the inequality 
 becomes
\bea
 & &  \phi_{F_1^E}^2(0) P^2(0) F_1^{2E}(0)
 +  \phi_{H_V^E}^2(0) P^2(0) H_V^{2E}(0)
+ ( {\phi_{F_1^E}} P {F_1^E})_z^{\prime 2}(0)\nonumber\\
& + & ( {\phi_{H_V^E}} P {H_V^E})_z^{\prime 2}(0) 
+ ( {\phi_{F_1^E}} P {F_1^E})_z^{\prime \prime 2}(0)
 +  ( {\phi_{H_V^E}} P {H_V^E})_z^{\prime \prime 2}(0) \leq 1.
\eea
In this case the domain of the allowed values for the 
charge radius and the convexity is the interior of an ellipse 
in the $\rho^2-c$
 plane of the form
\be
(\rho^2 -{\bar \rho}^2)^2 + S[(c -{\bar c}) - 
T (\rho^2 -{\bar \rho}^2)]^2=K^2,
\ee
where ${\bar \rho}^2$, ${\bar c}$, $S$, $T$ and $K$ are readily determined 
 through a comparison with eq. (3.23).  

\sxn{RESULTS}

In this section we discuss the numerical results of our analysis.
 The lower and upper bounds on the charge radius $\rho^2$ is obtained from 
 Eq.(3.19) with basically reliable  inputs. Actually, most of the 
predicted values
 of the charge radius from various models are 
well below the upper bound $\rho^2
 \leq 4.5$. The lower bound $\rho^2 \geq -1.9 $ is not interesting, since from
 Bjorken's sum rule \cite{you}, the positivity of 
the charge radius is expected.
 On the other hand, Eq.(3.24) gives the correlation between the charge radius 
 $\rho^2$ and
 the convexity parameter $c$. In Eq.(3.24), the numerical values of the
 parameters are:
\be
{\bar \rho}^2=1.31,~S=256,~ {\bar c}=1.45,~T=1.56~~and~~ K=3.22.
\ee
With these parameters, the solid curve in Fig. 1 restricts the allowed range
 for the charge radius and the convexity parameter of the IW function
to lie   within the interior of an ellipse in the $\rho^2-c$ plane.

 This result can be used to test the compatibility of  some phenomenological
  models
for the IW function. Each  model 
leads to  definite values for  the 
 charge radius and the convexity. For illustration, we consider the
 following models.

(1) The {\bf MIT Bag model} \cite{mit} provides 
the form of IW function as
 \be
\xi(\omega)=({2\over {\omega + 1}})^{3.5 +{1.2\over \omega}}.
\ee
This model gives the values of charge radius $\rho^2$  and  
convexity parameter
 $c$  as $ 2.35$ and $3.95$ respectively.

(2)  In {\bf Simple Quark Model} \cite{sqm}, 
the IW function is well approximated by the
 formula
\be
\xi(\omega)=({2\over {\omega + 1}})^{1.32 +{0.70\over \omega}},
\ee
which predicts the computed value of the charge radius 
and the convexity parameter
 to be $1.0$ and $1.11$ respectively. 

(3)  {\bf QCD Sum Rule} \cite{qcdsr} predicts
\be
\xi(\omega)=({2\over {\omega + 1}})^{1\over 2}
{\rm exp}[- 0.8{{\omega -1}\over 
 {\omega + 1}}].
\ee
The value of the charge radius $\rho^2=0.65$ predicted in this model, 
 gives the corresponding value of the convexity to be $0.47$.

(4)  The IW function in the {\bf Skyrme  model}  
\cite{soliton} has the form
 \be
\xi(\omega)= 0.99 {\rm exp}[-1.3 (\omega -1)],
\ee
with  $\rho^2=1.3$ and $c=0.85$.

(5) The {\bf Relativistic Three Quark 
Model} \cite{rqm} calculation gives the approximate form of
 IW function near the zero recoil point to be
\be
\xi(\omega)=({2\over {\omega + 1}})^{1.7 +{1\over \omega}},
\ee
 which indicates the numerical values  
 $\rho^2 = 1.35$  and $c= 1.75$.

(6) In {\bf Infinite Momentum Frame Quark Model} 
 \cite{imf}, the overlap integrals 
for the hadronic wave functions of parent
 and daughter baryon  calculated in the infinite momentum frame leads 
 to the
following form of the IW function
\be
\xi(\omega)= ({1\over \omega})^{{2 m + 1}/2} 
{\rm exp}[-\kappa^2 {{\omega -1}\over {2 \omega}}]
{{ H_{2m}(\kappa {\sqrt{{\omega +1}\over {2 \omega}}})} 
\over {H_{2m}(\kappa)}},
\ee
 where
\be
H_l(x)=\int_{-x}^{\infty}d y (y+x)^l e^{-z^2},
\ee 
with $m=1$ and $\kappa=1.5$.
This gives the corresponding values of $\rho^2$ and $c$ to be  $3.04$ and 
$6.81$ respectively.

In Fig. 1, the corresponding 
values of charge radius and convexity parameter
 in these models are plotted as points.
 In Table I, the predicted value of $\rho^2$
  and the computed value of $c$ parameter  
 in these phenomenological models
 are shown. The fourth column of the table gives the allowed range of $c$
 for the predicted value of 
$\rho^2$ for each model to test the compatibility of
 the corresponding model with the dispersive bound.

 The predicted values of the charge radius and convexity parameter in 
the Simple Quark  Model  and QCD sum rule  
lie well within the allowed domain, provided
 by dispersive approach, the others lie close to 
the periphery or quite outside  
 this region. 
\newpage
\sxn{CONCLUSION}
In this paper we have related the 
 two-point function in QCD 
 corresponding to the  $\Lambda_b{\bar \Lambda_b}$
 pair production (evaluated at one loop)  to 
the $\Lambda_b$ elastic formfactors  via 
 analyticity, crossing symmetry
 and dispersion relations. Some techniques of complex analysis , i,e.
 conformal mapping, introduction of Blaschke factor and the 
Schur-Caratheodory type inequality 
are employed  to derive constraints on a  simple parametrization of
 the elastic formfactors. In the heavy quark limit, these elastic 
formfactors and the formfactors for the semileptonic 
decay $\Lambda_b \to \Lambda_c l{\bar \nu}$
 are connected to 
 a single  IW function through heavy quark symmetry. As a result, 
we have obtained model independent  dispersive bounds
 on the charge radius and the curvature of the 
IW function in an expansion around 
 the zero recoil point $\omega=1$.  
 These bounds turn out to have nontrivial implications 
on the compatibility of
a number of models, which we have depicted in Table I. 
The constraints can be made more stringent 
with the inclusion of  higher derivatives , which
would give stronger correlation  between  the charge radius and the 
 convexity parameter of the 
IW function by reducing the allowed domain 
 in $\rho^2 - c $ plane.

 In previous works \cite{boyd4, boyd5}, 
this dispersive approach  has been  applied directly to
the formfactors for the $\Lambda_b \to \Lambda_c l {\bar \nu}$ 
decay. But the  power in that approach was  
somewhat limited  due to the possible presence 
of $b{\bar c}$ bound states below the 
threshold for $\Lambda_b{\bar \Lambda_c}$
production, which have not been observed experimentally
and thus require inputs
from potential models. Our approach on the other hand 
involves  only the elastic formfactors and  the 
use  of experimentally known $b{\bar b}$ bound states ($\Upsilon$s)
enables us to impose more stringent bounds on the IW function without
having to rely much on specific models. This approach can also be 
employed to study heavy baryon lifetimes and such works are in 
progress.

\newpage
\centerline{\large\bf Figure Captions}

\begin{enumerate}
\item The charge radius $\rho^2 $ and
 the convexity parameter $ c$
 are constrained to lie within the ellipse shown in figure. 
 Values of $\rho^2 <0$ are excluded by Bjorken's Sum Rule.
 The points in the figure denote the numerical values of the charge
 radius and the convexity, predicted in different phenomenological models.
 The numbers appearing next to the points in the figure correspond to the
 models indicated by bold faced letters in Section 4.
\end{enumerate}

\newpage
\begin{center}
\vskip .5cm
TABLE I\\
\vskip .5cm
\begin{tabular}{|c|c|c|c|}\hline
 & & & \\
Model &  Value &  Value & Allowed range of \\ 
& of $\rho^2$ & of $c$ & $c$ in $\rho^2-c$ plane\\
& & &  \\ \hline
& & &  \\
MIT bag Model &  2.35 & 3.95 & 2.89 - 3.27\\
& & &  \\ \hline
& & &  \\
Simple Quark Model & 1.01 & 1.11 & 0.79-1.19\\
& & &  \\ \hline
& & &  \\
QCD Sum Rule & 0.65 & 0.47 & 0.23-0.63\\
& & &  \\ \hline
& & &  \\
Skyrme Model &  1.30 & 0.85 & 1.25 - 1.64\\
& & &  \\ \hline
& & &  \\
Relativistic Three-Quark Model & 1.35 & 1.75 & 1.32-1.72\\
& & &  \\ \hline
& & &  \\
Infinite Momentum Frame Quark Model & 3.04 &6.81 & 3.98-4.32\\
& & &  \\ \hline
\end{tabular} \\
\end{center}
\vskip .5cm
\vskip .5cm

\centerline{\psfig{figure=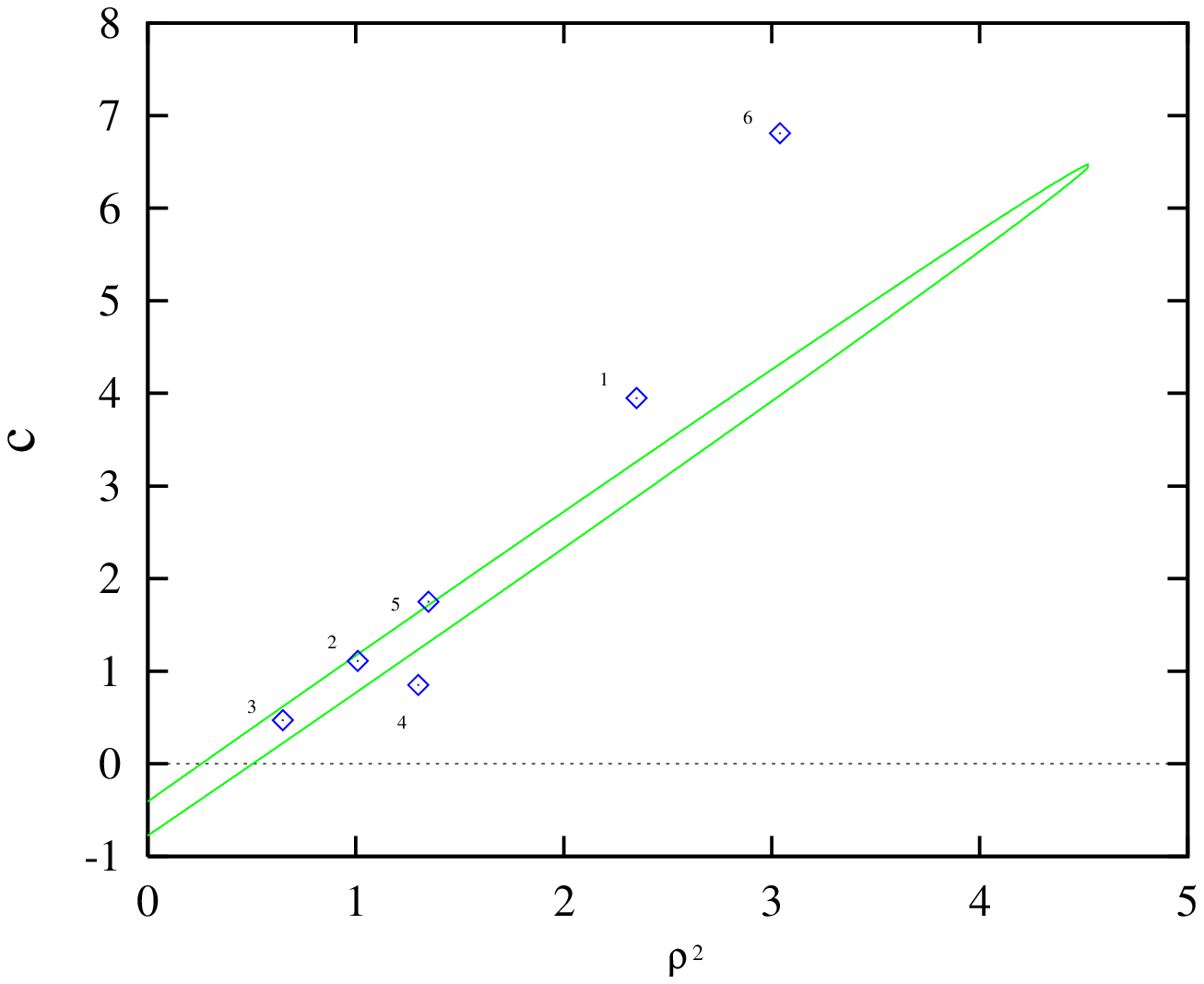,width=18.5cm,height=18.5cm}}
\vskip 1.0cm
 
\bigskip
\end{document}